\documentclass[submission,copyright,creativecommons]{eptcs}
\providecommand{\event}{SCAV 2018} % Name of the event you are submitting to
\usepackage{breakurl} % Not needed if you use pdflatex only.
\usepackage{underscore} % Only needed if you use pdflatex.
\usepackage {graphicx}
\usepackage {amsmath}
\usepackage {amssymb}
\title{Robust Safety for Autonomous Vehicles through Reconfigurable Networking}
\author{*Khalid Halba\qquad\qquad *Charif Mahmoudi \qquad\qquad **Edward Griffor
\institute{National Institute of Standards and Technology Gaithersburg, Maryland, USA}
\institute{* Advanced Network Technologies Division / Information Technology Laboratory}
\institute{** Smart Grid and Cyber-Physical Systems Program Office / Engineering Laboratory }
\email{\quad khalid.halba@nist.gov \quad\qquad charif.mahmoudi@nist.gov\quad\qquad edward.griffor@nist.gov}
}
\def\titlerunning{Robust Safety for Autonomous Vehicles through Reconfigurable Networking}
\def\authorrunning{K. Halba, C. Mahmoudi \& E. Griffor}
\begin{document}
\maketitle

\begin{abstract}
Autonomous vehicles bring the promise of enhancing the consumer's experience in terms of comfort and convenience and, in particular, the safety of the autonomous vehicle. Safety functions in autonomous vehicles such as Automatic Emergency Braking and Lane Centering Assist rely on computation, information sharing, and the timely actuation of the safety functions. One opportunity to achieve robust autonomous vehicle safety is by enhancing the robustness of in-vehicle networking architectures that support built-in resiliency mechanisms. Software Defined Networking (SDN) is an advanced networking paradigm that allows fine-grained manipulation of routing tables and routing engines and the implementation of complex features such as failover, which is a mechanism of protecting in-vehicle networks from failure, and in which a standby link automatically takes over once the main link fails. In this paper, we leverage SDN network programmability features to enable resiliency in the autonomous vehicle realm. We demonstrate that a Software Defined In-Vehicle Networking (SDIVN) does not add overhead compared to Legacy In-Vehicle Networks (LIVNs) under non-failure conditions and we highlight its superiority in the case of a link failure and its timely delivery of messages. We verify the proposed architecture’s benefits using a simulation environment that we have developed and we validate our design choices through testing and simulations.
\end{abstract}

\section{Introduction and Related Work}

\begin{figure}
\centering
\resizebox{0.6\columnwidth}{!}{\includegraphics{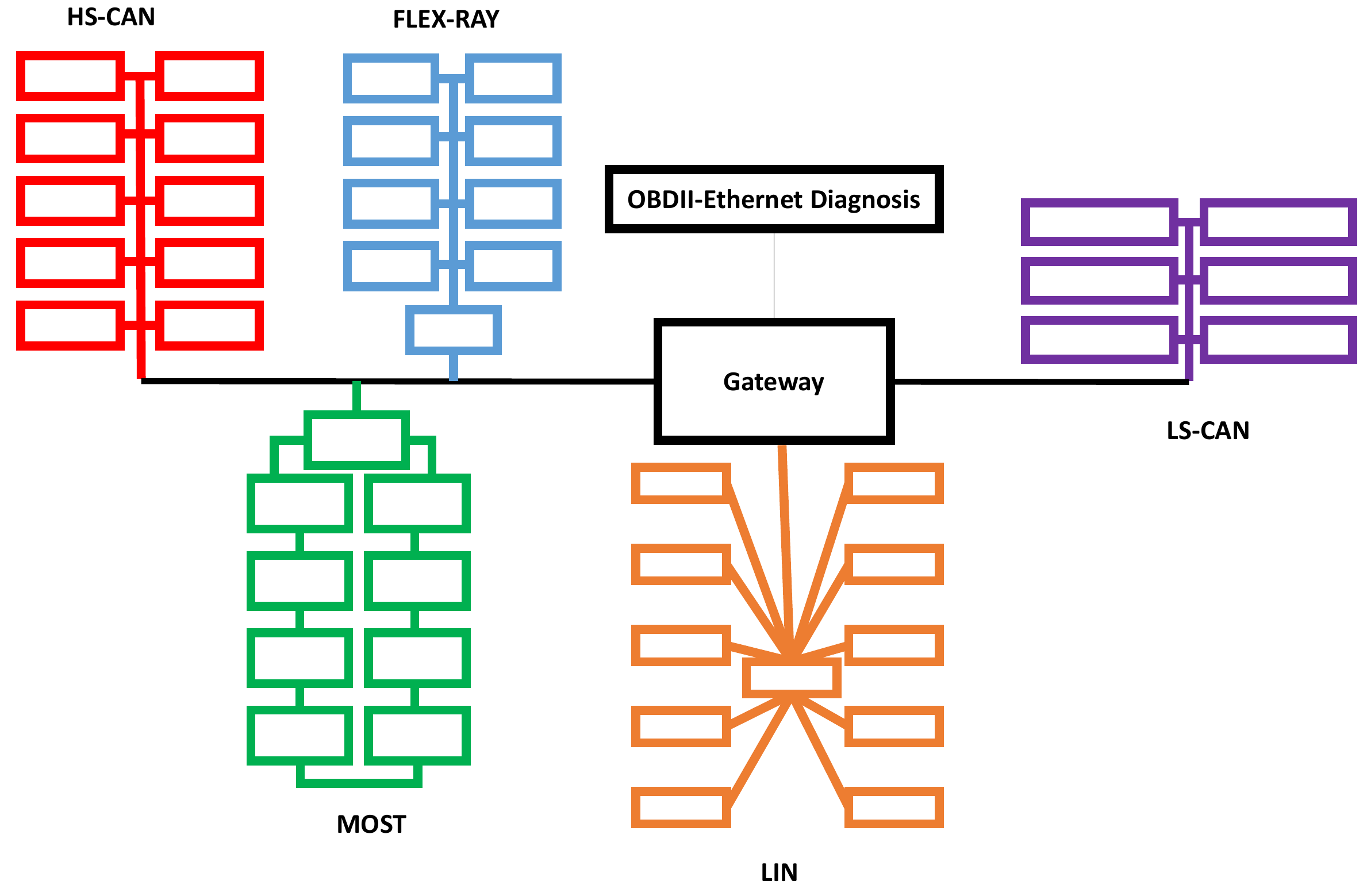}}
\caption{Legacy in-vehicle networks (LIVN) architecture}
\label{fig:LIVN}
\end{figure}

Safety is a key concern in the development of autonomous vehicles \cite{fernandes2012platooning}. In fact, according to IBM \cite{ghosh2007self}, autonomous cars by 2025 are expected to be equipped with self-healing mechanisms that enable decreased human intervention and maintenance. Self-healing features such as network reconfiguration in the case of a link failure are not natively supported in LIVNs \cite{tuohy2015intra}. Attempts to add support for failover mechanisms in LIVNs such as the Controller Area Network (CAN) \cite{philips} will come with the price of extra complexity and a hefty development cost. A damaged in-vehicle bus or Electronic Control Unit (ECU) can cause functional exclusion and the consequences might be catastrophic when it comes to safety-critical features. 

Introducing complex features such as failover is a design and cost challenge: Car manufacturers tend to introduce features that leverage existing components and design. This is not only a cost saving and risk minimization approach, but also saves the manufacturers significant research and development cost and time. With the advances in network technologies such as SDN and Time Triggered Ethernet (TTE), the introduction of new designs that support complex features without changing the underlying LIVN ECUs \cite{Steinbach2011realteth} \cite{du2016software} is now possible. The usual solution when existing LIVN buses cannot support new features is to re-architect, which introduces new risk and cost throughout the product line engineering. 

Resiliency in LIVNs has been the subject of research efforts from industry and academia. In \cite{philips}, Philips analyzed the different use cases associated with CAN bus failures and proposed Redundant Transmission Technique to overcome link or transceiver failure. This technique relies simply on the replication of the ECU's components as well as the CAN bus itself. This allows the CAN system to use a second transceiver or bus if the first one fails. Applying the same technique to the other LIVNs such as LIN (Local Interconnect Network: a single wire network mainly used for non-critical purposes such as body control), Flex Ray (considered as the successor of the CAN bus as it handles more traffic and advanced control features), or MOST(Media Oriented Systems Transport : an optical bus designed to carry infotainment traffic) will lead to an increase in cost, weight, and environmental impact, all of which steer decision making in the automotive industry. Another study highlighted the risk associated with losing parts or most of an in-vehicle bus\cite{lima2016towards}: The authors categorized the attacks on the in-vehicle bus as a threat to the integrity and the availability of the vehicle's features. A native threat to the resiliency of LIVNs, especially the CAN bus, is its data transmission paradigm that relies on broadcasting data between components \cite{tuohy2015intra}. The fact that a CAN frame is received by all the ECUs constitutes a threat to the in-vehicle bus since attackers need only access to one ECU to be able to eavesdrop on all the messages circulating in the bus or inject false messages that could change the behavior of the vehicle. TTE represents a flavor of deterministic and real time Ethernet technologies. It was tested and proved a viable solution for automotive \cite{steinbach2010comparing} and aeronautics \cite{loveless2015ttethernet} applications. TTE switches also supports fault tolerance \cite{rushby2001bus}, which makes it a perfect candidate for future autonomous vehicle's network backbone. SDN is a networking paradigm that leverages separation of control and data plane and network programmability to enable complex features such as fine grained forwarding \cite{bianco2010openflow}, QoS \cite{tomovic2014sdn}, load balancing \cite{belyaev2014towards}, and failover \cite{beheshti2012fast}.

As opposed to the LIVN architecture described in Fig. \ref{fig:LIVN}, the architecture we propose enables Fast Failover \cite{beheshti2012fast} in the case of a link failure and message delivery in a timely manner regardless of data frequency or refresh rate, this will solve the lack of failover support by LIVNs. Our design has also the added benefit of leveraging unicast communications instead of broadcasts, reducing the risk of eavesdropping on automotive components \cite{lima2016towards}. Our contribution features SDN as an overlay on top of the CAN bus and the other LIVNs. This overlay will enable complex features that can stimulate innovation and help realize IBM's 2025 self-healing car vision.

In this work, we propose a SDIVN (defined above) with support of Fast Failover to enable autonomous vehicles self-healing in the case of a link failure. This design can scale to support more links and ECUs, which will improve resilience, sustainability and interoperability. We discuss our contributions as follows: In Section 2 we discuss the elements of the design we propose, in Section 3 we leverage the design elements to build a simulation environment for validation. In Section 4, we stress the simulation environment through experimentation, results generation, and results discussion. Finally, we summarize the work we have done in this paper and we highlight current and future work in Section 5.

\section{SDIVN Design Elements}

\begin{figure}
\centering
\resizebox{0.6\columnwidth}{!}{\includegraphics{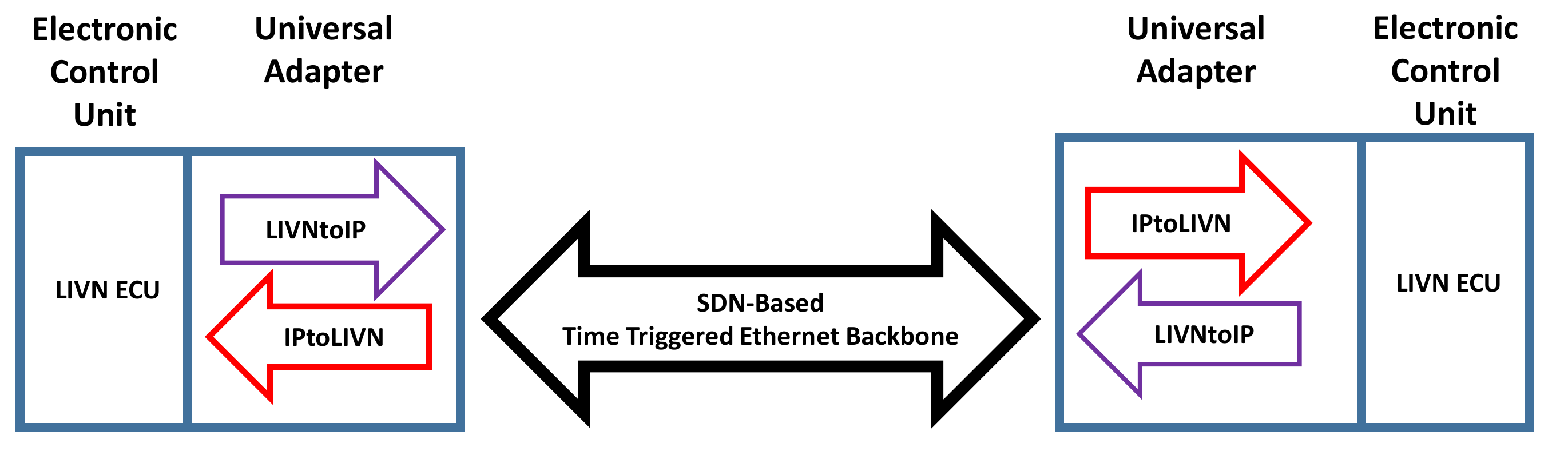}}
\caption{Each ECU is composed of an LIVN message Generator/Receiver and a Universal Adapter that Encapsulates/De-encapsulates ECUs generated/received messages}
\label{fig:ADAPTER}
\end{figure}
\begin{figure}
\centering
\resizebox{1\columnwidth}{!}{\includegraphics{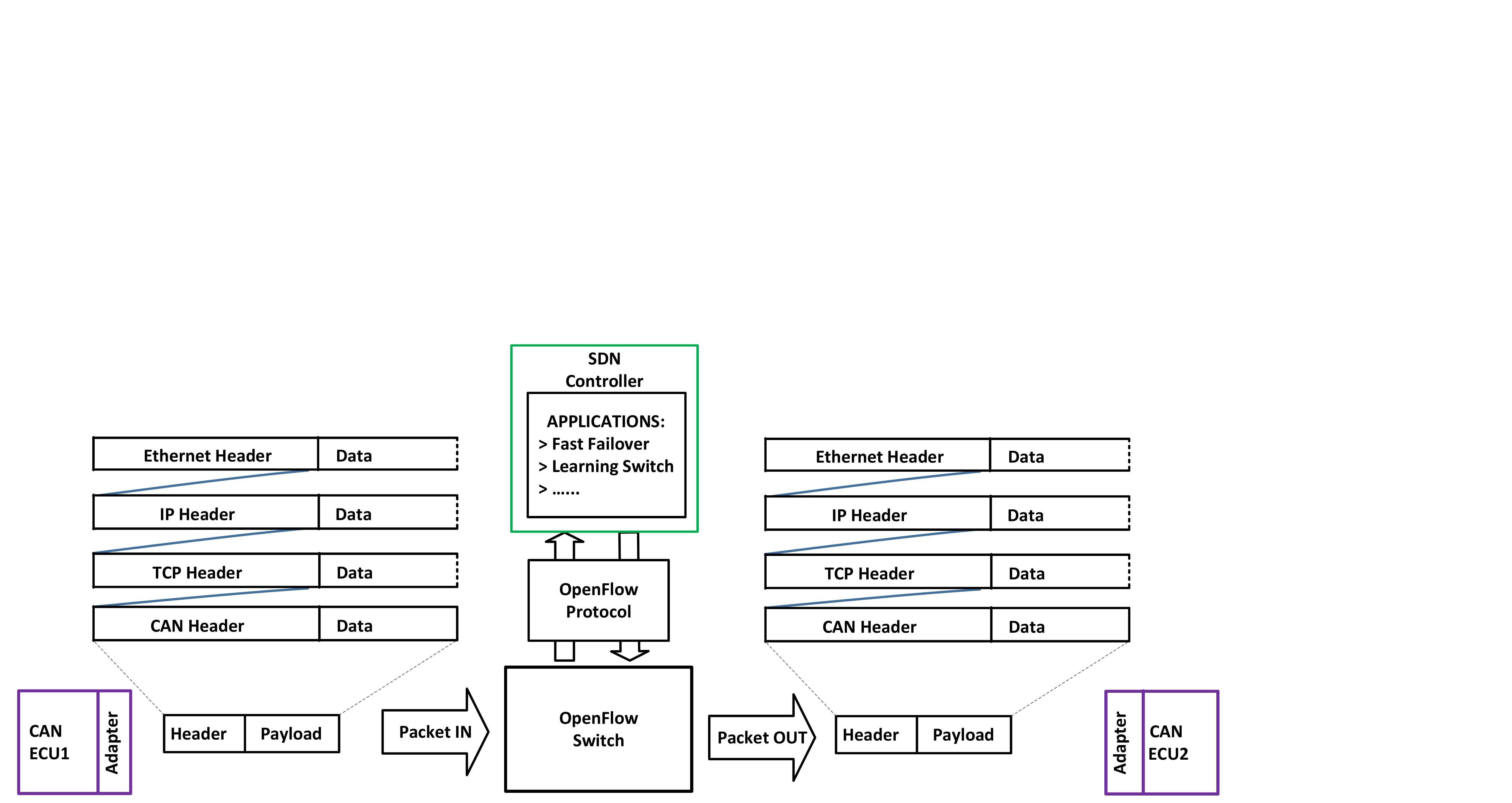}}
\caption{Software Defined In-Vehicle Network design and the journey of a CAN ECU message}
\label{fig:SDIVN}
\end{figure}

In this section we present the SDIVN design that enables the implementation of advanced features within an automotive network. This design not only assures safety mechanisms but also interoperability between different ECU protocols and features. In this work we focus on the safety mechanisms and we leave the interoperability aspect for a future work. The SDIVN is composed of the following components: Electronic Control Units (ECUs) generate and receive data from off/on-board sensor arrays or from other ECUs. These ECUs are IP capable thanks to the universal adapters described in Fig. \ref{fig:ADAPTER} that enable the communication across the SDN backbone by re-packing LIVN messages (such as CAN messages) to an Ethernet frames and vice-versa. These ECUs are connected to a TTE Backbone through Open Flow capable switches \cite{limoncelli2012openflow}. The choice of TTE as a backbone is due to the fact that it outperforms industrial automotive technologies such as Flex Ray \cite{Steinbach2011realteth} in terms of real time requirements, flexibility, and resiliency. The Open Flow switches are connected to the SDN controller which also runs additional applications related to, for example, resiliency \cite{da2015resilience}, load balancing \cite{belyaev2014towards}, QoS \cite{yates2017managing}, security \cite{dacier2017security}, etc.

\begin{figure}
\centering
\resizebox{0.5\columnwidth}{!}{\includegraphics{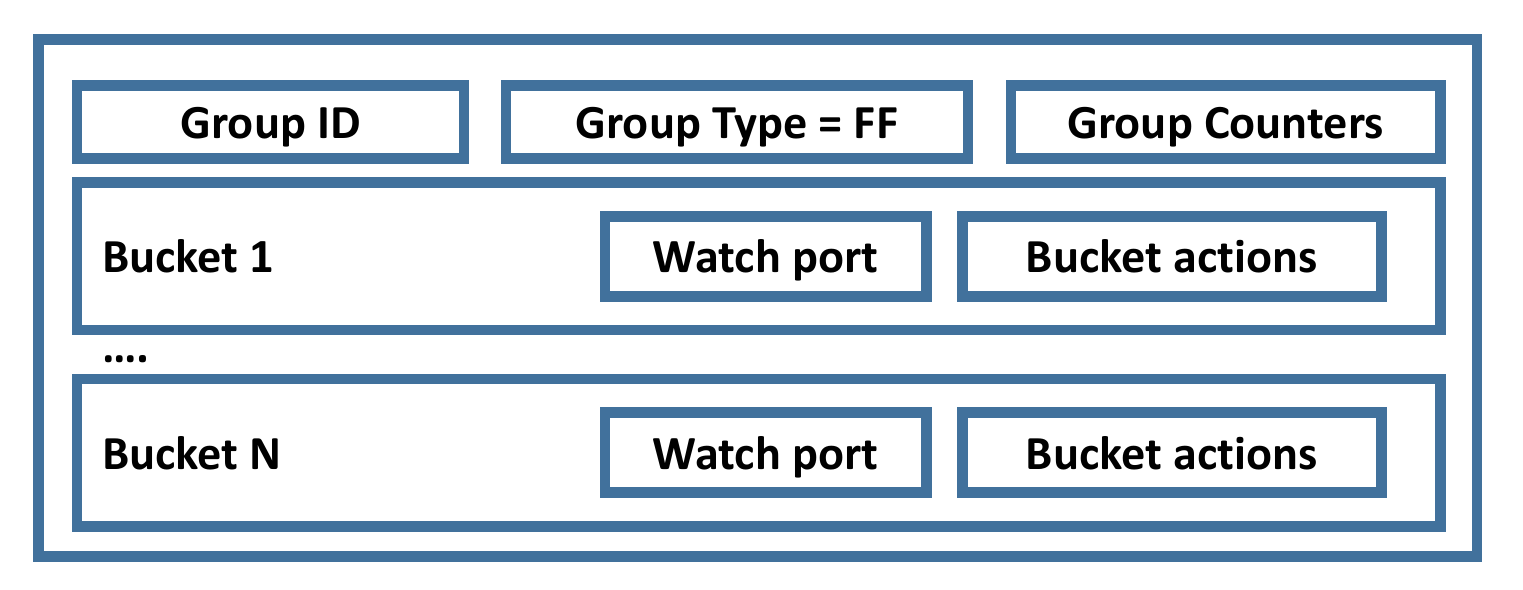}}
\caption{ Fast Failover Group Elements}
\label{fig:FFelements}
\end{figure}

The process of generating data by an ECU includes re-packing it into an Ethernet frame by the adapter, processing it through the SDN backbone, subjecting it to the Open Flow rules within SDN switches, and delivering it to the destination ECUs. This process is described in Fig. \ref{fig:SDIVN}.

The final element in our design is the Fast Failover application that runs in the SDN controller. Fast Failover \cite{beheshti2012fast} is a feature that allows link reconfiguration in the case of a port failure. This feature leverages Open Flow 1.3 Groups to enable port state monitoring and action upon port status change, using watch ports and action buckets. The SDN application that runs at the SDN controller populates SDN switches with flow tables and rules that help the network recover from link failure. Rules include Fast Failover Groups that implement mechanisms of path switching if a link is down. Fig. \ref{fig:FFelements} is an illustration of Fast Failover Group components, the components of relevance are the buckets that contain a port status monitoring watch port; If the latter fails packets are processed according to actions defined in the bucket actions and are eventually sent to backup paths. More details about the dynamics of Fast Failover are explained in \cite{lin2016fast}. Tests have been run to assess this feature and results show that the time to recover is in the order of microseconds which is acceptable for the applications we intend to test.

\section{Design Validation}

In this section, we describe the resiliency validation design for both SDIVN and LIVN. An obstacle detection use case is leveraged to highlight how each design operates regarding the resiliency feature.

Obstacle detection in autonomous cars is a key stone property that requires reliable and complete perception of the environment of the vehicle in order to avoid high risk situations. Numerous works have investigated obstacle detection techniques and requirements in autonomous vehicles: In \cite{pietsch1998autonomous} the authors analyzed sonar image data properties in autonomous underwater vehicles and introduced a special sonar data compression technique that optimizes the obstacle avoidance process and enhances sonar data interpretation. In \cite{bernini2014real} the authors presented a survey on obstacle detection techniques that are based on stereo vision or 2D/3D technologies. For our work, we are rather interested in preserving critical functions such as obstacle detection during a bus failure by leveraging the SDIVN network programmability capability. 

For the SDIVN and LIVN the components behavior is described in Fig. \ref{fig:valSDIVN} as follows: CAN1 ECU and CAN3 ECU generate different CAN messages with different frequencies, each message represents a specific automotive function. For illustration purposes, CAN3 represents a special hardware that detects obstacles in front of the car and sends messages to the Antilock Braking system (ABS) located at CAN4 at a high refresh rate of 1 message per 50 ms. The ABS system ensures the vehicle's stopping based on road conditions and information that comes from the obstacle detection ECU located at CAN3 ECU. CAN1 represents a sensor that measures the road condition and sends messages to an alarm system located at CAN2 ECU at a low refresh rate of a 1 message per 100ms. Frequencies used here may be different for different applications, they are used here for illustrating the use of SDIVN. We want to verify that under normal conditions (no link failure) the design guarantees data delivery and also ensures the integrity of the signals frequencies at the receiving ECUs. We also want to establish the benefit of the Fast Failover mechanism when a link fails.

\subsection{SDIVN}

\begin{figure}
\centering
\resizebox{0.8\columnwidth}{!}{\includegraphics{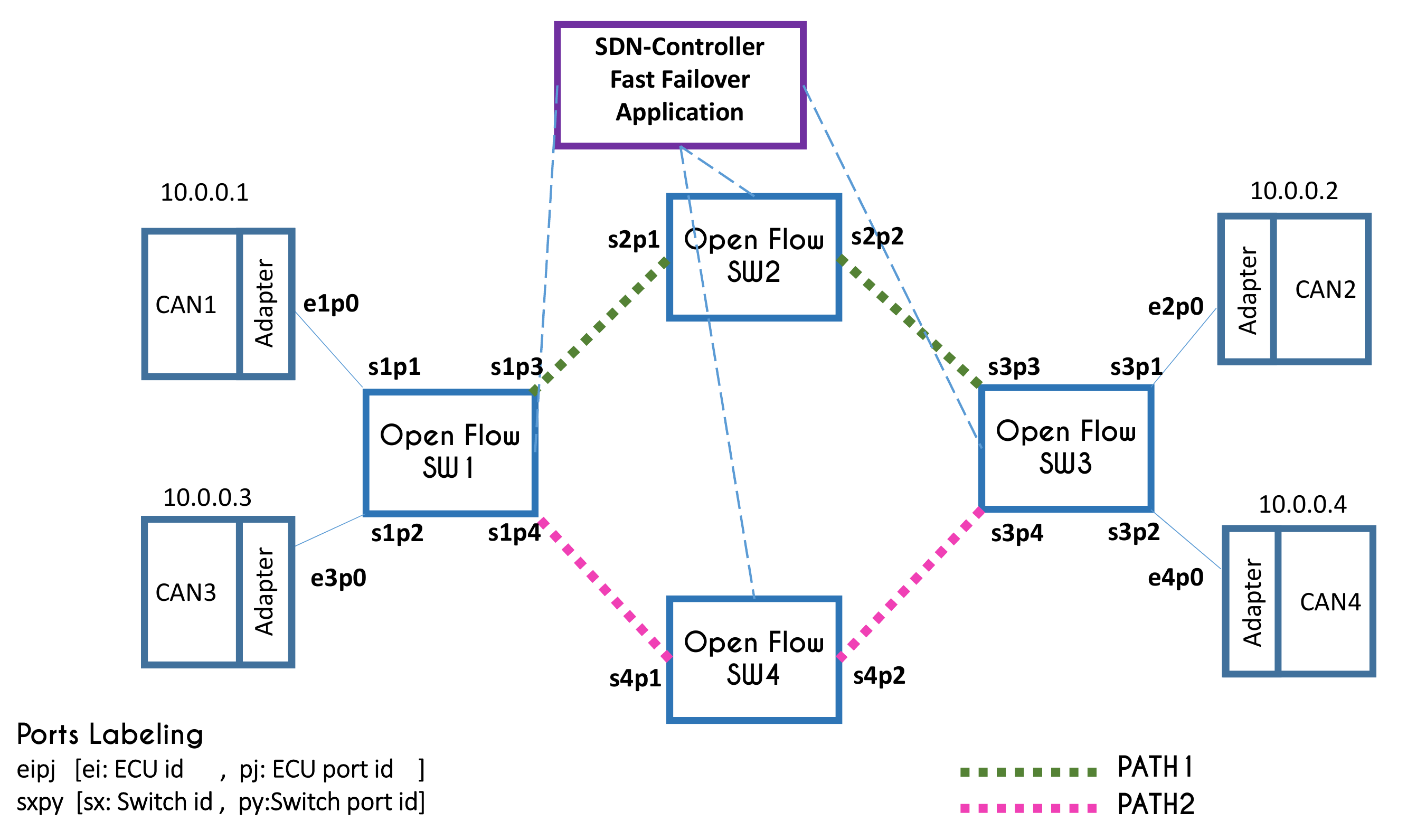}}
\caption{ ECUs exchange in-vehicle messages and leverage the SDIVN}
\label{fig:valSDIVN}
\end{figure}

We validate the proposed Software Defined In-Vehicle Network for link failover by the implementation of the different components described in the previous section and running a simulation of the proposed design. Fig. \ref{fig:valSDIVN} illustrates the validation use case for the SDIVN. 

We are aware that the size and design of the SDIVN in Fig. \ref{fig:valSDIVN} is not representative of the wide spectrum of possible sizes and architectures. We are reducing the scope in purpose in order to make a proof of concept for the resiliency property.   

We care about the integrity of the message frequency because as we approach an obstacle, if the frequency is altered at the reception, the detection or the expected actuation process might take longer than it should, and appropriate reaction to it in a timely manner might not be possible. 

\begin{figure}
\centering
\resizebox{1\columnwidth}{!}{\includegraphics{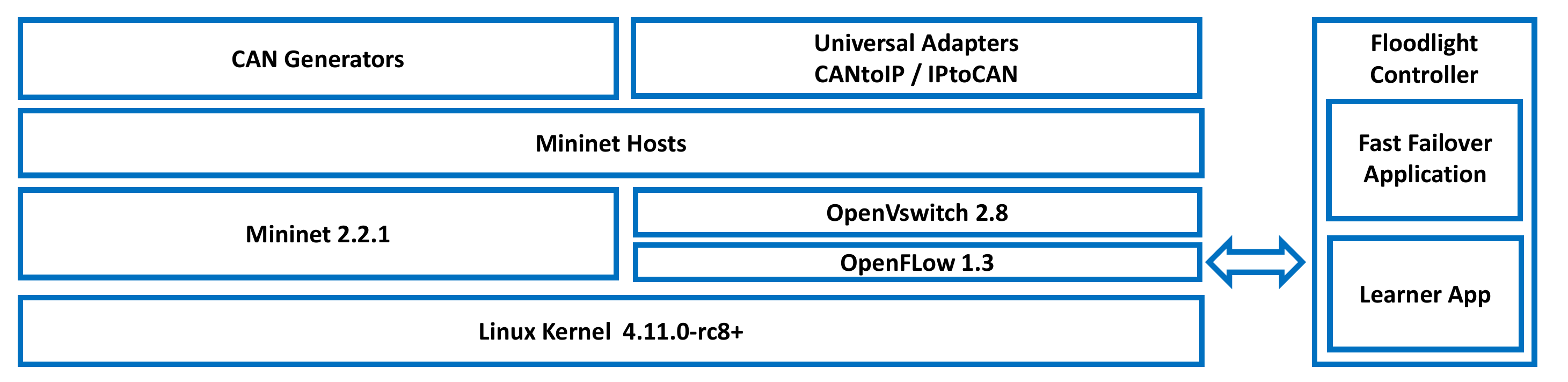}}
\caption{ Developed and Open-Source components used in the validation experiment}
\label{fig:stackSDIVN}
\end{figure}

We have implemented the simulation environment using Mininet 2.2.1 installed on a patched Linux Kernel (4.11.0-rc8+) that supports CAN-Utils Package namespaces [14]. Additionally, we have built the protocol adapters based on TCP/IP Sockets (that ensure retransmission in the case of a packet loss) and CAN-Sockets \cite{CAN-Utils}. For the SDN implementation, Floodlight \cite{shalimov2013advanced} is our choice for the controller as it supports the implementation of Fast Failover. A representation of the software stack we have leveraged is described in Fig. \ref{fig:stackSDIVN}

\begin{figure}
\centering
\resizebox{0.6\columnwidth}{!}{\includegraphics{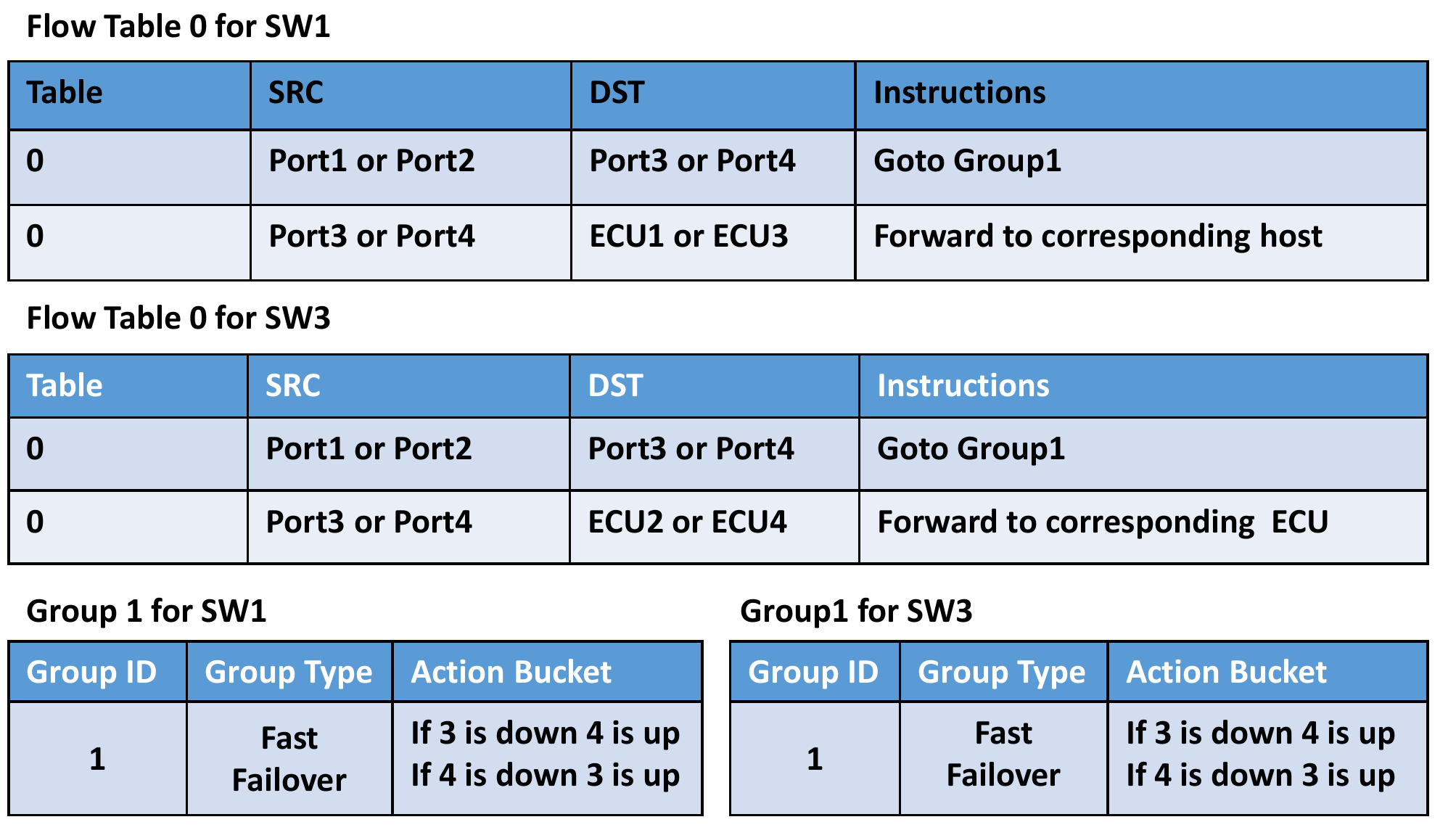}}
\caption{ Open Flow tables and Group tables after the SDN controller runs the Fast Failover application}
\label{fig:flow table}
\end{figure}

After running the simulation, the SDN controller populates the flow and group tables with necessary rules and actions to handle link failover Fig. \ref{fig:flow table} is a non-exhaustive perspective on the group table and flow table rules installed after running the application. The hosts are discoverable using the learning switch application that runs in the beginning of the process. The fast failover application operates as follows: One path can be used at a time, if PATH1 is enabled, PATH2 ports are disabled and vice versa, and ports from different paths can't be enabled at the same time. If there is a port failure on one path, the Fast Forwarding application disables ALL ports on that path and switches to another path (see Fig. \ref{fig:valSDIVN} for illustration).The code for the application we have used for this setup is available in \cite{CAN2IP}.

\subsection{LIVN}
\begin{figure}
\centering
\resizebox{0.5\columnwidth}{!}{\includegraphics{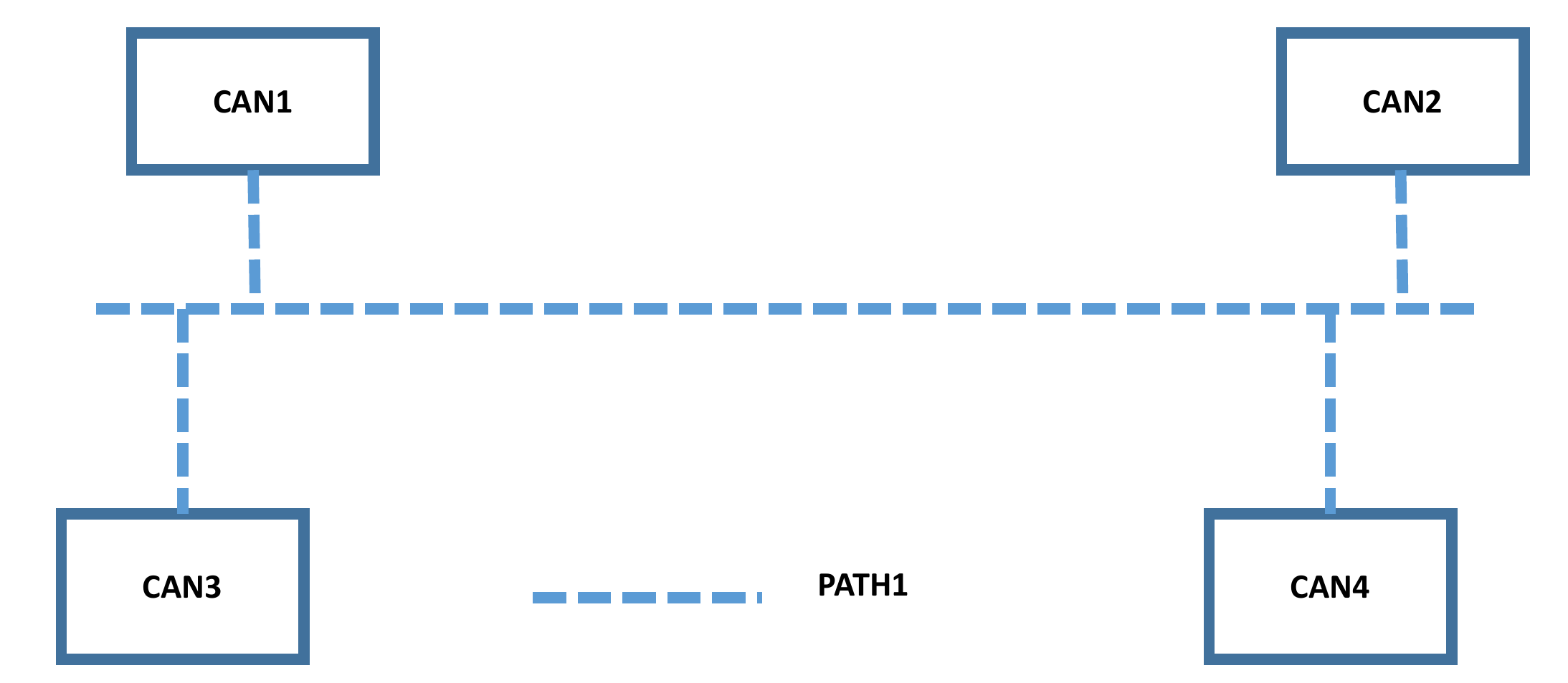}}
\caption{ECUs exchange CAN messages and leverage a broadcast medium to simulate a CAN bus}
\label{fig:valCANBUS}
\end{figure}

For the CAN bus simulation, we have adopted the design in Fig. \ref{fig:valCANBUS}. This design can't support multiple paths for failover as the CAN bus is a broadcast medium and adding extra links to it will result in broadcast storms, resulting in replicate messages in the beginning of the data transfer, and a complete outage of the bus after few seconds. These are some of the limitations of the CAN bus. 

\section{Results Discussion}

In this section we show the benefit of the Fast Failover resiliency mechanism in the SDIVN design we have proposed, and we estimate the Average Failover Cost per packet which will inform us about whether or not the impact of the Fast Failover mechanism is significant.

\subsection{Results outcome for SDIVN and LIVN}

\begin{figure}
\centering
\resizebox{0.8\columnwidth}{!}{\includegraphics{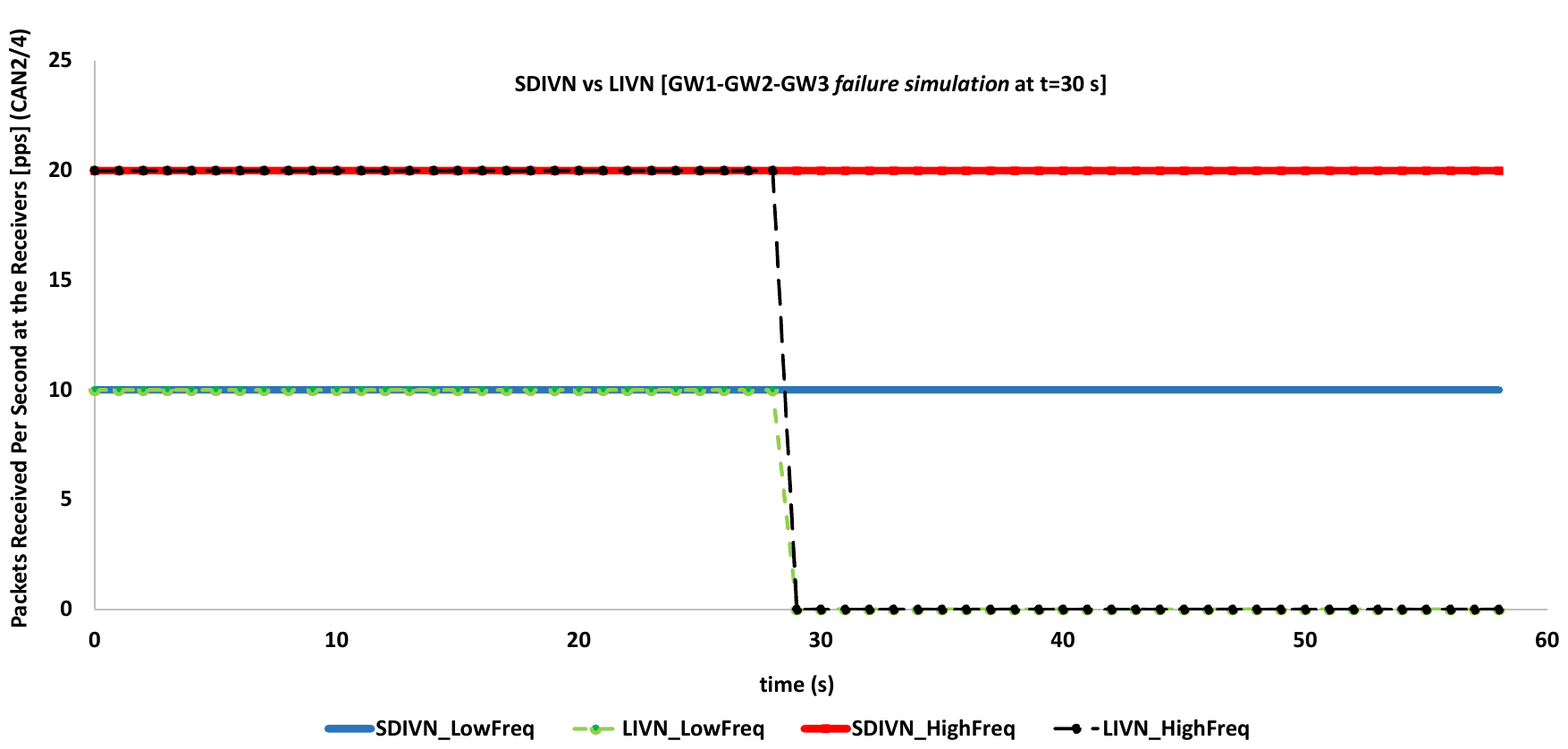}}
\caption{Fast Failover results in both cases: SDIVN instantly switches to PATH2 when PATH1 fails. LIVN fails to maintain function since it does not have failover mechanisms}
\label{fig:FastFailoverResults}
\end{figure}

We have run experiments described in the previous section for 60 seconds, and we simulate a link failure at t=30 seconds. Heuristic experiments show that the transition from one failed path to an up and running path in the case of SDIVN is almost instantaneous which is illustrated in Fig. \ref{fig:FastFailoverResults}. The figure shows how the SDIVN design we proposed preserved the function and its properties while the LIVN didn't. SDIVN allows the network to quickly recover from link failure and preserves message delivery and frequency. Another benefit of the SDIVN is its unicast nature, messages are sent to their specific destinations based on rules generated by the application running in the SDN controller, reducing the risk of bus outage or the replicate message caused by broadcasts. In the same figure we see that the LIVN is incapable of recovering from a link failure as it does not support resiliency mechanisms.

\subsection{Timing Analysis} 

The main focus of this subsection is to estimate the delay cost of the Fast Failover mechanism. For this end, we describe the timing parameters that are involved when a packet is processed through the proposed SDIVN.\newline

We define the total propagation delay $T_pd$ as the sum of the propagation delays of the N links $P(i)$ connecting sending and receiving endpoints.

\begin{equation}
\label{eq:capacity}
T_pd=\sum_{i=1}^{N}P(i)
\end{equation}

We define the Transfer Time $T_p$ for $1$ packet from a sender to a receiver as the difference in time between the reception time $T_r$ of the packet at the receiving ECU and the sending time  $T_s$ at the generating ECU, as expressed in (\ref{eq:T1}). 

\begin{equation}
\label{eq:T1}
T_p = T_r - T_s \newline
\end{equation}

$T_p$  can also be expressed as follows (\ref{eq:T2})\newline

\begin{equation}
\label{eq:T2}
T_p = T_pd + ED_d + F_d + C_d + FO_d
\end{equation}

Where $T_pd$ is the total propagation delay along the path between the two ECUs, $ED_d$ is the time it takes to encapsulate and decapsulate the packet at the adapter, $F_d$ is the forwarding time at the Open Flow switches, $C_d$ is the processing time at the SDN controller, and $FO_d$ is the failover time in the case of a link failure\newline

The encapsulation time and decapsulation time $ED_d$ is the time it takes the universal adapter at the ECU's edge to Encapsulate $E_d$ and Decapsulate $D_d$ the packet as expressed in (\ref{eq:T3})
\begin{equation}
\label{eq:T3}
ED_d = E_d + D_d.
\end{equation}

The failover time $FO_d$ is the time it takes the system to detect failure $DF_d$, reconfigure the network $RN_d$, and retransmit the packet $RP_d$ if it was lost during failure, as expressed in (\ref{eq:T4})

\begin{equation}
\label{eq:T4}
FO_d = DF_d + RN_d + RP_d
\end{equation}

We define the Transfer Time for all packets $TT_d$ as the sum of the Transfer Time of all received packets $T_p(i)$. K is the total number of received packets, as expressed in (\ref{eq:T5})

\begin{equation}
\label{eq:T5}
TT_d =\sum_{i=1}^{K}T_p(i)
\end{equation}

We define the average packet transfer time ATT as follows (\ref{eq:T6})

\begin{equation}
\label{eq:T6}
ATT =TT_d/K.
\end{equation}

Under normal conditions (no link failure), $ATT$ = $ATTN_p$, and if a link failure occurs during the transmission process, $ATT$=$ATTF_p$. The relationship between $ATTF_p$ and $ATTN_p$ is expressed in (\ref{eq:T7})

\begin{equation}
\label{eq:T7}
AFCP_p = ATTF_p - ATTN_p
\end{equation}
\begin{figure}
\centering
\resizebox{0.4\columnwidth}{!}{\includegraphics{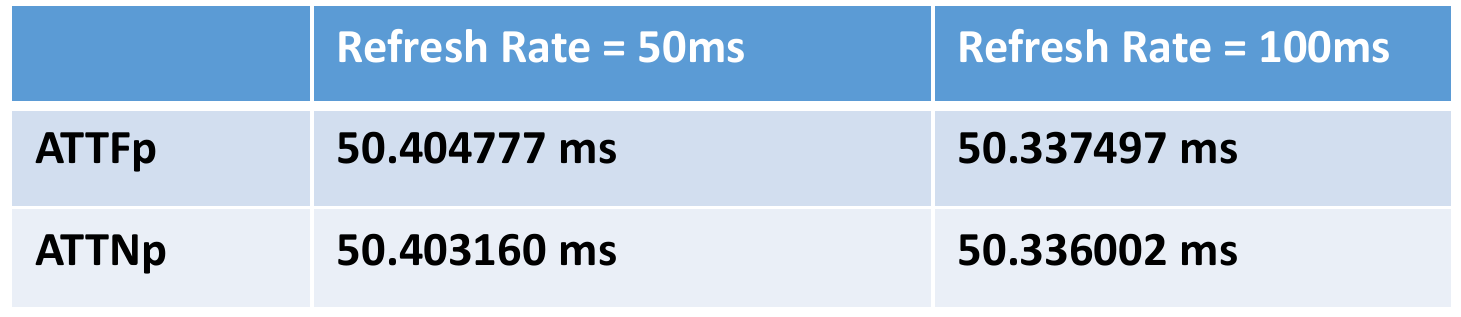}}
\caption{$ATTN_p$ and $ATTF_p$ for both frequencies.}
\label{fig:ATT}
\end{figure}

For a single link failover we calculate the Average Failover Cost Per Packet $AFCP_p$ for both refresh rates (50 ms and 100 ms). We can express $AFCP_p$ as the difference in time between the Average Packet Transfer Time during normal conditions $ATTN_p$ and the Average Packet Transfer Time during link failover $ATTF_p$. This calculation will give us insights on whether or not Fast Failover has a negative impact on the functionality. \newline

We calculate the $AFCP_p$ based on the $ATTF_p$ and $ATTN_p$ values we have measured and logged in Fig. \ref{fig:ATT} 

\paragraph{$AFCPp$[50 ms]} = $ATTFp$ - $ATTNp$ = 1.617 us = 0.00320\% of the $ATTNp$[50 ms].
\paragraph{$AFCPp$[100 ms]} = $ATTFp$ - $ATTNp$ = 1.495 us = 0.00297\% of the $ATTNp$[100 ms].

\subsection{Summary of Results} 

The Average Failover Cost Per Packet $AFCPp$ value in both cases (50 ms and 100 ms refresh rate) is negligible compared to the average transfer time per packet $ATTN$. We conclude that the Fast Failover mechanism does not cause a significant delay overhead when a link failure occurs in this example. However, it guarantees timely message delivery and message frequency integrity. 

\section{Conclusion}

In this work, we presented a new design for in-vehicle networks that allows recovery from link failure by leveraging the network programmability provided by SDN. Since automotive networks are size determined, engineers could use SDIVN to design failover paths for a functionality that they judge to be critical. We have demonstrated through simulation that this mechanism is so efficient that the time to recover is insignificant for the frequencies and network size we have tested, and we believe that the positive results perceived during experiments we ran will hold for larger networks. In-Vehicle network components and links will continue to evolve in terms of hardware performance and software optimisation to accomodate the bandwidth and delay requirements.

For Future work, we are exploring other interesting areas such as the impact of interoperability between different ECUs. We believe that interoperability between heterogeneous data sources within an SDIVN will stimulate the innovation of a new spectrum of advanced autonomous vehicle features. We are also conducting a statistical analysis based on the fractional factorial design \cite{gunst2009fractional}, the goal is to understand which parameters (network size, bus speed, bus degradation, bus delay ...) have the most influence on certain autonomous vehicle properties of interest(resiliency, interoperability, scalability ...). Another area of interest is Software Defined Wireless In-Vehicle Networks (SDWIVN), for which we built a testbed and conducted promising experiments \cite{WirelessSDIVN}, such design choice has the potential of bringing novel contributions to energy saving techniques and environmental impact improvement.

\section*{NIST Disclaimer}

Any mention of commercial products or organizations is for informational purposes only; it is not intended to imply recommendation or endorsement by the National Institute of Standards and Technology, nor is it intended to imply that the products identified are necessarily the best available for the purpose. The identification of any commercial product or trade name does not imply endorsement or recommendation by the National Institute of Standards and Technology, nor is it intended to imply that the materials or equipment identified are necessarily the best available for the purpose. Certain commercial entities, equipment, or materials may be identified in this document in order to describe an experimental procedure or concept adequately. Such identification is not intended to imply recommendation or endorsement by NIST, nor is it intended to imply that the entities, materials, or equipment are necessarily the best available for the purpose.

\nocite{*}
\bibliographystyle{eptcs}
\bibliography{generic}

\end{document}